\documentclass[epj,onecolumn]{webofc}

\usepackage[varg]{txfonts}
\usepackage{microtype}
\usepackage{bm}

\newcommand{\dd}{\mathrm{d}}
\newcommand{\as}{\alpha_{\mathrm{s}}}
\newcommand{\eps}{\epsilon}
\newcommand{\Ti}{\bm{T}_{i}}
\newcommand{\Tj}{\bm{T}_{j}}
\newcommand{\order}{\mathcal{O}}

\def\beq{\begin{equation}}
\def\eeq{\end{equation}}
\def\beqa{\begin{eqnarray}}
\def\eeqa{\end{eqnarray}}
\def\eq#1{Eq.~(\ref{#1})}

\begin{document}

%%%%%%%%%%%%%%%%%%%%%%%%%%%%%%%%%%%%%%

\title{A celestial viewpoint on infrared divergences in QCD}

\author{
\firstname{Lorenzo} \lastname{Magnea}\inst{1}\fnsep
\thanks{\email{lorenzo.magnea@unito.it}}
\and
\firstname{Enrico} \lastname{Zunino}\inst{2}}

\institute{Dipartimento di Fisica, Universit\`a di Torino, 
and INFN, Sezione di Torino, \\
Via Pietro Giuria 1, I-10125 Torino, Italy
\and
Higgs Centre for Theoretical Physics, School of Physics and Astronomy, \\
The University of Edinburgh, Edinburgh EH9 3FD, Scotland, United Kingdom}

\abstract{%
Non-abelian infrared radiation can be seen as both a practical ingredient for QCD 
phenomenology and a window on the long-distance structure of gauge theories. During
the last decade, a novel viewpoint was developed on infrared phenomena: soft currents
are seen as arising from asymptotic symmetries realised on the {\it celestial sphere}, 
and soft colour correlations similarly emerge from a two-dimensional {\it conformal 
field theory} on the sphere. In this contribution, we briefly review the status of this 
approach. We begin with the standard bulk picture, presenting the infrared factorisation 
of massless scattering amplitudes, and then we show how universal infrared information 
gets mapped to marked points on the celestial sphere. In particular, soft colour-dipole
singularities are reproduced to all orders by a correlator of vertex operators in a 
free-boson theory on the sphere. We then proceed to study soft-gluon currents, and 
show how bulk results constrain the structure of the celestial theory beyond the 
free-field limit. For single emission, we note that all logarithms in dipole-like loop 
contributions can be absorbed into the choice of scale for the strong coupling. 
For multiple emission, we show how the emerging mixed-helicity celestial operator product 
expansion must be weighted by energy fractions, breaking holomorphic factorisation 
and obstructing associativity under successive collinear limits. These results illustrate 
a remarkable connection between the familiar physics of infrared QCD and the dynamics 
of a two-dimensional theory on the celestial sphere.
}

%%%%%%%%%%%%%%%%%%%%%%%%%%%%%%%%%%%%%%

\maketitle

%%%%%%%%%%%%%%%%%%%%%%%%%%%%%%%%%%%%%%%%%%%%%%%%%%%%%%%%%%%%%%%%%%%%%%%

\section{Introduction}
\label{sec:introduction}

Infrared (IR) singularities are encountered in almost every facet of precision QCD calculations.  
To briefly illustrate their origin, consider a massless parton of momentum $p$, emitting a 
gluon of energy $\omega_k$ at an angle $\theta_{pk}$. The Feynman diagram for this process
will feature a propagator denominator of the form
\beq
\label{eq:soft-collinear-origin}
 (p+k)^2 \, = \, 2 p \cdot k \, = \, 2 E_p \omega_k (1 - \cos\theta_{pk}) \, .
\eeq
This denominator will vanish (and thus the particle will go on shell) when the gluon is soft, 
$\omega_k \to 0$, or collinear to the quark, $\theta_{pk} \to 0$. The corresponding singularities 
arise from long-distance configurations, since on-shell particles can propagate indefinitely: 
soft or collinear emission cannot resolve the short-distance scattering and is not suppressed 
at late times.  At fixed perturbative order, these configurations produce logarithmic phase-space 
integrals and virtual poles in dimensional regularisation.  More fundamentally, they display the 
fact that isolated coloured Fock states are not the true asymptotic states of an unbroken gauge 
theory.

As is well known, physically meaningful cross sections are nevertheless finite. The 
Kinoshita--Lee--Nauenberg (KLN) theorem indeed guarantees that summing over experimentally 
indistinguishable real and virtual configurations will result in the cancellation of their 
soft and collinear singularities~\cite{Kinoshita:1962ur,Lee:1964is}. For QCD phenomenology, 
this statement is, however, a starting point, rather than the end of the problem. Real and 
virtual corrections live in different phase spaces, and the cancellation of singularities
must be implemented analytically, through some form of subtraction or phase-space slicing, 
before a stable numerical integration becomes possible. Moreover, restrictions imposed by a 
measurement often lead to an imperfect cancellation, leaving behind logarithms of ratios of 
scales.  When, for example, a resolution scale $Q_0$ is much smaller than a hard scale $Q$, 
these logarithms can compensate the smallness of the coupling through terms of the form
$\as^n\log^m(Q/Q_0)$, which must then be resummed. This is possible thanks to the universality
of infrared radiation, which leads to the factorisation of scattering amplitudes, separating 
hard, collinear and wide-angle soft dynamics. Factorisation, in turn, implies evolution 
equations, whose solutions lead to resummation.  The same universality underlies threshold 
resummation, transverse-momentum resummation, infrared subtraction, and the formalism of 
parton showers: for a broad review, see Ref.~\cite{Agarwal:2023}.

For soft radiation, at leading power, universality means that a soft gluon is sensitive only 
to the direction and color charge of hard partons, not to the details of the hard interaction 
that created it.  Hard partons can therefore be replaced by semi-infinite Wilson lines following
their classical trajectories.  Products of such lines encode wide-angle soft exchange, while
collinear radiation can be assigned to individual external legs.  The renormalisation of Wilson-line 
operators leads to a matrix evolution equation in color space, and explains the exponentiation of 
IR poles. A central ingredient of exponentiation is the light-like cusp anomalous dimension: 
physically, it measures radiation associated with a massless colour charge changing direction, 
and it is ubiquitous throughout infrared QCD studies.

The celestial approach organises this familiar long-distance picture in a geometric and
holographic setup. Every null Wilson line starting from the interaction point ends at a 
point on the sphere at null infinity: the hard scattering is thus encoded in a configuration 
of points on the punctured sphere. Lorentz transformations act on that sphere as $S\!L(2,\mathbb C)$,
which is the group of global conformal transformations 
in two Euclidean dimensions. Infrared radiation is then also encoded in a pattern imprinted
on the celestial sphere. This intuition begins to build Strominger's {\it infrared triangle},
relating three descriptions of the same IR physics: soft theorems, asymptotic symmetries, 
and memory effects \cite{Strominger:2014,Strominger:2017}.  In the gauge-theory realisation, 
asymptotic symmetries are angle-dependent gauge transformations that remain non-trivial at 
null infinity, while memory is the lasting imprint left by soft radiation on the colour 
structure of the hard scattering; the leading-order soft theorem becomes the Ward identity 
of the corresponding current on the sphere. The most radical realisation of this picture
builds {\it celestial amplitudes}, by taking a Mellin transform of ordinary amplitudes, and 
replacing particle energies with boost weights, quantum numbers conjugate to logarithmic energy 
rescalings~\cite{Arkani-Hamed:2020gyp}.  Here we take a deliberately narrower route, retaining 
the ordinary QCD factorisation, and attempting to build a celestial theory only for the
infrared factor of the amplitude, which we regard as naturally holographic.

An important test of this idea was provided in Ref.~\cite{Magnea:2021}, where it was shown 
that every colour-dipole contribution to the IR factor of fixed-angle amplitudes can be reproduced, 
to all perturbative orders, by a correlator of matrix-valued vertex operators in a free-boson 
theory on the sphere.  The construction is non-trivial, despite the word ``free'': non-commuting 
colour generators act on the hard amplitude, and the complete dependence on the coupling, regulator 
and renormalisation scale enters through a matching coefficient built from the cusp anomalous 
dimension. On the other hand, higher colour multipoles and real-emission loop corrections are 
not generated by this Gaussian theory, so it must be understood to give a precise baseline 
rather than a complete holographic version of the infrared sector of non-abelian theories.

In the next two sections, we will briefly review the infrared factorisation of bulk gauge theory 
amplitudes, and the dipole construction of Ref.~\cite{Magnea:2021}, in a language intended to 
connect directly with standard QCD practice.  With that foundation in place, we will turn, 
in the last section, to more recent work, focusing on the constraints imposed on the celestial 
theory by our understanding of real soft radiation beyond leading order.

%%%%%%%%%%%%%%%%%%%%%%%%%%%%%%%%%%%%%%%%%%%%%%%%%%%%%%%%%%%%%%%%%%%%%%%

\section{Infrared factorisation of scattering amplitudes}
\label{sec:factorisation}

Consider a renormalised amplitude for $n$ massless hard particles with momenta $p_i$, in 
a {\it fixed-angle} configuration, where all Mandelstam invariants $s_{ij} \equiv 2 p_i 
\cdot p_j$ are parametrically of the same size. In dimensional regularisation, with 
$d=4-2\eps$ and $\eps<0$ for infrared singularities, the following factorisation formula 
holds~\cite{Agarwal:2023}:
\beq
\label{eq:amplitude-factorisation}
  \mathcal A_n \left(\frac{p_i}{\mu}, \as(\mu), \eps \right) \, = \, 
  Z_n \left( \frac{p_i}{\mu}, \as(\mu), \eps \right) \,
  \mathcal H_n \left( \frac{p_i}{\mu}, \as(\mu), \eps \right) \, .
\eeq
Here $\mathcal H_n$ and $\mathcal A_n$ are vectors in the space of color configurations for the
amplitude at hand, while $Z_n$ is a color operator acting on $\mathcal H_n$. The hard factor 
$\mathcal H_n$ is finite as $\eps \to 0$, while $Z_n$ generates all soft and collinear poles.  
The finite hard function $\mathcal H_n$ depends on the short-distance process, whereas $Z_n$ 
depends only on the species, momenta and color representations of the external coloured particles.  
This is the sense in which IR singularities are universal. The scale evolution of $Z_n$ is 
controlled by a finite, matrix-valued anomalous dimension, according to
\beq
\label{eq:Z-solution}
  Z_n \, = \, P \exp \left[ \frac12 \int_0^{\mu^2} \frac{\dd \lambda^2}{\lambda^2} \, 
  \Gamma_n \left(\frac{p_i}{\lambda}, \as(\lambda,\eps) \right) \right] \, ,
\eeq
where the path ordering $P$ is needed because colour matrices at different scales need not
commute.  An anomalous dimension is the kernel governing how a renormalised operator changes 
with scale: here it is a colour matrix because soft exchanges can rotate colour configurations,
even as they don't affect hard momenta at leading power.  Although $\Gamma_n$ is finite, the 
lower endpoint of the scale integral is potentially singular, and indeed the integration of
the $d$-dimensional running coupling down to vanishing scale is the source of {\it all} infrared 
poles.  

In order to represent explicitly the IR factor $Z_n$, a common notation introduces colour 
insertion operators $\bm T_i^a$, which act as gauge generators in the color representation 
of leg $i$, but are realised as the identity on all other legs. Consequently, they obey
\beq
\label{eq:colour-algebra}
  \big[ \bm T_i^a, \bm T_j^b \big] \, = \, 
  \mathrm{i}f^{abc} \bm T_i^c \, \delta_{ij} \, ,
  \qquad
  \bm T_i^2 \, = \, C_i \, ,
  \qquad
  \sum_{i=1}^{n} \bm T_i \, = \, 0 \, ,
\eeq
where $C_i$ is the quadratic Casimir eigenvalue for leg $i$. The last equation implements colour 
conservation, and applies when the operators act on a gauge-invariant hard amplitude: it is the 
non-abelian analogue of charge neutrality, and it will arise as a zero-mode constraint in the 
calculation of the celestial correlator.

Remarkably, the general structure of the massless infrared anomalous dimension $\Gamma_n$ is well 
understood. Up to terms involving higher-order Casimir invariants, which arise starting at four loops, 
$\Gamma_n$ separates into a color-dipole contribution, and a multipole correction depending only on 
conformal cross-ratios of hard momenta, which starts contributing at three loops:
\beq
\label{eq:gamma-decomposition}
  \Gamma_n \, = \, \Gamma_n^{\, \mathrm{dip}} (s_{ij}, \as) + 
  \Delta_n \big( \rho_{ijkl}, \as \big) \, ,
  \qquad
  \rho_{ijkl} \, \equiv \, \frac{(p_i \! \cdot p_j)(p_k \! \cdot p_l)}{(p_i \! \cdot p_k)
  (p_j \! \cdot p_l)} \, .
\eeq
The color-dipole contribution is known to all orders~\cite{Gardi:2009,Becher:2009}, and is given by
\beq
\label{eq:dipole-gamma}
  \Gamma_n^{\mathrm{dip}} (s_{ij}, \as) \, = \, \frac12 \, \widehat \gamma_K(\as)
  \sum_{i<j} \log \left(\frac{-s_{ij}+\mathrm{i}\eta}{\mu^2} \right) \, 
  \Ti \cdot \Tj - \sum_i \gamma_i(\as) \, ,
\eeq
where $\widehat\gamma_K$ is the light-like cusp anomalous dimension with the quadratic Casimir 
removed, and $\gamma_i$ is a color-diagonal collinear anomalous dimension. The scale invariance 
of semi-infinite Wilson lines plays a pivotal role to strongly constrain \eq{eq:dipole-gamma}
\cite{Gardi:2009,Becher:2009}.  Genuine quadrupole correlations first occur at three loops through 
$\Delta_n$, and are explicitly known at that order in terms of weight-five single-valued harmonic 
polylogarithms~\cite{Almelid:2016}.

%%%%%%%%%%%%%%%%%%%%%%%%%%%%%%%%%%%%%%%%%%%%%%%%%%%%%%%%%%%%%%%%%%%%%%%

\section{A celestial construction for infrared dipole correlations}
\label{sec:free-boson}

The starting point of the construction in Ref.~\cite{Magnea:2021} is the colour-correlated part 
of the bulk operator $Z_n$, which can be obtained by stripping off hard collinear effects. Since 
this factor contains only long-distance information, and depends on hard partons only through 
their light-like directions, it is natural to ask whether it can be computed by a theory living 
directly on the sphere which represents the locus of those directions.  The answer is remarkably 
simple for dipole color correlations. A key ingredient is the celestial parametrisation of 
light-like momenta, to which we now turn.

%%%%%%%%%%%%%%%%%%%%%%%%%%%%%%%%%%%%%%

\subsection{Parametrising the infrared factor on the celestial sphere}
\label{subsec:coordinates}

With metric signature $(+---)$, a future-pointing massless momentum can be parametrised by
giving an energy and a point on the Riemann sphere, according to
\beq
\label{eq:celestial-momentum}
  p_i^\mu \, = \, \frac{\omega_i}{\sqrt2} \, \Bigl(1 + z_i \bar z_i, \ z_i + \bar z_i, \
  - \mathrm{i}(z_i - \bar z_i), \ 1 - z_i \bar z_i \Bigr) \,.
\eeq
Writing $z_{ij} = z_i-z_j$, and similarly for $\bar z_{ij}$, one obtains
\beq
\label{eq:celestial-products}
  p_i \cdot p_j \, = \, \omega_i \omega_j |z_{ij}|^2 \, ,
  \qquad \quad
  \beta_i^\mu \, \equiv \, p_i^\mu/\omega_i
  \quad \rightarrow \quad
  \beta_i \cdot \beta_j \, = \, |z_{ij}|^2 \, .
\eeq
Thus, Mandelstam invariants factor into energies and distances between marked points.
Furthermore, the cross-ratios in \eq{eq:gamma-decomposition} become
\beq
  \rho_{ijkl} \, = \, |z_{ijkl}|^2 \, ,
  \qquad
  z_{ijkl} \, = \, \frac{z_{ij}z_{kl}}{z_{ik}z_{jl}} \, .
 \label{eq:cross-ratio}
\eeq
Cross-ratios are invariant under global M\"obius transformations; intuitively, they represent 
the shape variable characterising four marked points, after the three complex degrees of freedom 
of $SL(2,\mathbb C)$ have been removed.  One sees that the allowed non-dipole kinematics is already
expressed in the natural language of a two-dimensional conformal theory.

In the dipole sector, and in the parametrisation of \eq{eq:celestial-momentum}, it is easy to 
see that colour conservation can be used to separate the energy logarithms from the correlated 
angular dependence~\cite{Magnea:2021}.  Uncorrelated, colour-singlet terms, including all collinear 
poles, are local on the sphere and are not predicted by the proposed celestial correlator. 
The correlated factor is
\beq
 \label{eq:celestial-Z}
  Z_n^{\,\mathrm{corr}}(z_{ij},\as,\eps) \, = \, \exp \left[- K(\as,\eps) 
  \sum_{i<j} \log|z_{ij}|^2 \, \Ti \cdot \Tj\right] \, ,
\eeq
with
\beq
 \label{eq:K-definition}
  K(\as(\mu),\eps) \, = \, - \frac12 \int_0^\mu \frac{\dd\lambda}{\lambda} \, 
  \widehat \gamma_K \bigl( \as(\lambda,\eps) \bigr) \, .
\eeq
All scale, regulator and coupling dependence is in the single universal function $K(\as, \eps)$; 
all colour and angular dependence is in the sum over puncture pairs.  This complete separation is 
the key that makes an exact two-dimensional representation possible. The coefficient $K$ is built 
from the light-like cusp anomalous dimension, and is a well-known building block in perturbative 
QCD, appearing in form factors, threshold resummations and coherent parton branching, motivating 
its interpretation as a universal infrared effective coupling \cite{Catani:2019}. In high-energy
scattering, it is also closely related to the perturbative Regge trajectory. The celestial correlator 
to be constructed in the next section will therefore package an element of QCD that is universal 
across many different observables. 

%%%%%%%%%%%%%%%%%%%%%%%%%%%%%%%%%%%%%%

\subsection{A free-boson celestial representation of the dipole infrared factor}
\label{subsec:free-boson}

In order to derive \eq{eq:celestial-Z} from a celestial field theory, we introduce free scalar 
fields $\phi^a(z,\bar z)$ taking values in the Lie algebra of the gauge group, with action
\beq
\label{eq:free-action}
  S[\phi] \, = \, \frac{1}{2\pi} \int\dd^2z \,
  \partial_z \phi^a \, \partial_{\bar z} \phi_a \, .
\eeq
At the level of the action, the fields are uncorrelated; non-commuting colour is carried instead 
by vertex operators acting on the hard factor.  At every marked point on the sphere, we define a 
normal-ordered matrix-valued vertex operator of the form
\beq
\label{eq:vertex}
  V_i (z_i,\bar z_i) \, = \, :\!\exp \Bigl( \mathrm{i} \kappa \, \bm T_i \cdot
  \bm \phi(z_i, \bar z_i) \Bigr) \!: \, .
\eeq
We now construct a correlator of $n$ such vertex operators, mirroring the correlation of
the $n$ Wilson lines responsible for the soft poles of the amplitude. The correlator
can be evaluated exactly in the free theory, and, remarkably, one finds
\beq
\label{eq:correlator-match}
  \left\langle \prod_{i=1}^{n} V_i( z_i, \bar z_i) \right\rangle \, = \, 
  Z_n^{\mathrm{corr}}(z_{ij},\as,\eps) \, ,
\eeq
up to an overall colour-singlet normalisation. More precisely, the zero-mode integral 
enforces colour conservation, while Wick contraction of the non-zero modes gives 
the pairwise logarithms. Finally, the field normalisation $\kappa^2$ is matched to 
$K(\as,\eps)$. \eq{eq:correlator-match} is an all-order result, covering every contribution 
to $Z_n$ with dipole colour structure: it includes both planar and non-planar effects 
through the all-order cusp function in $K(\as,\eps)$.

Note that a generic matrix exponential would not behave like a primary field in a conformal 
theory. In this case however, colour conservation can be used to show that the vertex operators 
in \eq{eq:vertex} have a properly defined positive conformal weight. The correlator therefore 
has the expected conformal properties, even though it remains an operator in colour space.

Importantly, the free action also has holomorphic and anti-holomorphic Noether currents 
associated with translations in field space,
\beq
\label{eq:noether-currents}
  j^a(z) \, = \, \partial_z \phi^a(z,\bar z) \, ,
  \qquad \quad
  \bar j^a(\bar z) \, = \, \partial_{\bar z} \phi^a(z,\bar z) \, .
\eeq
The Ward identities for these currents read
\beq
\label{eq:ward-identity}
  \left\langle j^a(z) \prod_i V_i(z_i,\bar z_i) \right\rangle
  \, \propto \, \sum_i \, \frac{T_i^a}{z - z_i} \, 
  \left\langle \prod_i V_i(z_i,\bar z_i) \right\rangle \, ,
\eeq
with the complex-conjugate equation for $\bar j^a$. To compare this with QCD radiation, 
one can choose polarisation vectors
\beq
\label{eq:polarisation}
  \varepsilon_+^\mu(q) \, = \, \frac{1}{\sqrt2} \big( \bar z,1,- \mathrm{i},
  - \bar z \big) \, ,
  \qquad
  \varepsilon_-^\mu(q) \, = \, \frac{1}{\sqrt2} \big(z, 1, \mathrm{i}, -z \big) \, .
\eeq
In this parametrisation, the tree-level QCD currents for the emission of soft gluons with
positive (negative) helicity have the simple expressions
\beq
\label{eq:tree-soft-celestial}
  \omega_q S_a^{+,\, (0)}(q) \, = \, \sum_i \, \frac{T_i^a}{z_{qi}} \, ,
  \qquad \quad
  \omega_q S_a^{-,\, (0)}(q) \, = \, \sum_i \, \frac{T_i^a}{\bar z_{qi}} \, .
\eeq
Hence, \eq{eq:ward-identity} is precisely the positive-helicity soft-gluon theorem,
and the anti-holomorphic Ward identity gives its negative-helicity partner.  This 
connection underlies the interpretation of the tree-level soft theorem in terms
of a celestial current algebra of Kac--Moody type~\cite{He:2016}.

Notice that, while the correlator in \eq{eq:correlator-match} reproduces virtual dipoles 
to all orders, the free-theory Ward identity in \eq{eq:ward-identity} reproduces real soft
emission only at tree level. In fact, in QCD, loop corrections to the soft current are 
proportional to structure constants and originate from gluon self-interactions. They are 
direct evidence that the complete celestial theory, if it exists, must be interacting. 
Constraining these interactions by means of field theory data is the topic of the next 
section.

%%%%%%%%%%%%%%%%%%%%%%%%%%%%%%%%%%%%%%%%%%%%%%%%%%%%%%%%%%%%%%%%%%%%%%%

\section{Celestial constraints from high-order real soft radiation}
\label{sec:single-soft}

We will now present a summary of the recent analysis of Ref.~\cite{Magnea:2026}, which 
makes use of known high-order QCD soft currents as data pointing to the structure of the 
interacting celestial theory. In particular, we will highlight two conclusions. First, 
we note that logarithms arising beyond tree level in single-emission currents can be 
absorbed by a physical choice of the scale of the strong coupling. Second, we will 
derive from the structure of mixed-helicity multiple emission the requirement of an 
energy-weighted OPE for the celestial currents.

%%%%%%%%%%%%%%%%%%%%%%%%%%%%%%%%%%%%%%

\subsection{Single-soft logarithms and the physical dipole scale}
\label{subsec:one-loop}

Let us label hard partons by their celestial data, $\xi_i \equiv (\omega_i,z_i,\bar z_i)$.  
The leading-power factorisation for the soft emission of a gluon with momentum $q$ and
polarisation $\lambda$ can then be expressed as
\beq
\label{eq:real-factorisation}
  \mathcal A_{n+1}^\lambda \big( \xi_q; \{ \xi_i \} \big)
  \, = \, S_a^\lambda \big(\xi_q; \, \{z_i, \bar z_i \} \big) \, 
  \mathcal A_n \big(\{ \xi_i \} \big)\, ,
\eeq
where $S_a^\lambda$ is the soft factor. At one loop, the bare correction to the soft factor
is purely non-abelian. Rescaling invariance of Wilson lines, power counting and gauge 
invariance determine its kinematic form: only an $\eps$-dependent coefficient $C_1(\eps)$ 
requires integration \cite{Catani:2000}. Before making any scale choice, the one-loop bare 
current reads
\beqa
\label{eq:one-loop-general}
  \omega_q S_a^{\lambda, \, (1)}(z_q,\bar z_q) & \!\!\! = \!\!\! &
  \mathrm{i} C_1(\eps) \, f_{abc} \sum_{i \ne j} T_i^b T_j^c \, 
  \mathcal K_{ij}^{\lambda}(z_q,\bar z_q) \, 
  \left(
  \frac{\mu^2}{2 \omega_q^2} \, 
  \frac{|z_{ij}|^2}{|z_{qi}|^2|z_{qj}|^2}
  \right)^\eps \, , \nonumber\\
  \mathcal K_{ij}^{+} & \!\!\! = \!\!\! & \frac{z_{ij}}{z_{qi} z_{qj}} \, , \qquad \quad
  \mathcal K_{ij}^{-} \, = \, \frac{\bar z_{ij}}{\bar z_{qi} \bar z_{qj}} \, .
\eeqa
Thus, helicity selects a meromorphic or anti-meromorphic angular kernel $\mathcal K$, 
while the real factor raised to the power $\eps$ contains the dimensional scale dependence.
The explicit loop integration yields the prefactor
\beq
\label{eq:C1}
  C_1(\eps) \, = \, - \frac{e^{\eps\gamma_E}}{\eps^2} \,
  \frac{\Gamma^3 (1-\eps) \, \Gamma^2(1+\eps)}{\Gamma(1-2\eps)} \, .
\eeq
The last factor in \eq{eq:one-loop-general} assigns a different scale to every hard dipole.  
In celestial variables, the scale can be written as
\beq
\label{eq:qij-definition}
  q_{ij}^2 \, \equiv \, 2 \omega_q^2 \,
  \frac{|z_{qi}|^2|z_{qj}|^2}{|z_{ij}|^2}
  \quad \rightarrow \quad
  \left(
  \frac{\mu^2}{2 \omega_q^2}
  \frac{|z_{ij}|^2}{|z_{qi}|^2|z_{qj}|^2} \right)^\eps \, = \, 
  \left(\frac{\mu^2}{q_{ij}^2}\right)^\eps \, .
\eeq
In terms of Mandelstam variables, using \eq{eq:celestial-products}, one has $q_{ij}^2 =
s_{iq}s_{jq}/s_{ij}$. The physical meaning of $q_{ij}^2$ is well known: boosting to the 
centre-of-mass frame of particles $i$ and $j$ readily shows that $q_{ij}^2$ is the squared 
transverse momentum of the emitted gluon relative to dipole $(ij)$. We see that each pair 
of Wilson lines supplies its own physical radiation scale.

Next, we notice that the scale in \eq{eq:qij-definition} has exactly the power required 
to be absorbed into the $d$-dimensional running coupling, which obeys
\beq
\label{eq:d-dimensional-beta}
  \lambda \, \frac{\partial\as}{\partial\lambda} \, = \, - 2 \eps \as - 
  \frac{\as^2}{2\pi} \sum_{k \ge 0} \left(\frac{\as}{\pi}\right)^k b_k \, ,
\eeq
where $b_k$ are the coefficients of the $\beta$ function. At lowest order, the solution is 
simply $\as(q_{ij}^2,\eps)=(\mu^2/q_{ij}^2)^\eps \, \as(\mu^2,\eps)$. Consequently, the product 
of the coupling evaluated at scale $\mu$ with the final factor in~\eq{eq:one-loop-general} 
is the coupling evaluated at the dipole scale $q_{ij}$, to the perturbative accuracy 
required here. After this identification, the renormalised soft-emission factor has a 
simple and elegant form  in celestial variables. Dropping $\beta$-function contributions,
for positive helicity, one finds
\beq
\label{eq:one-loop-current}
  \frac{\as(\mu^2)}{4\pi} \, \omega_q \, S_a^{+, \, (1)}(q) \, = \, 
  \mathrm{i}C_1(\eps) \, f_{abc} \sum_{i \ne j} T_i^b T_j^c \, 
  \frac{z_{ij}}{z_{qi}z_{qj}} \, 
  \frac{\as(q_{ij}^2,\eps)}{4\pi} \, .
\eeq
The remaining rational factor $z_{ij}/(z_{qi}z_{qj})$ is dimensionless, scale invariant 
and meromorphic; for negative helicity, it is replaced by its complex conjugate.  

This observation changes the interpretation of the scale logarithms appearing in the 
one-loop current. Indeed, expanding $(\mu^2/q_{ij}^2)^\eps$ against the $1/\eps^2$ 
singularity in $C_1$ produces terms proportional to $\log q_{ij}^2$ and $\log^2 q_{ij}^2$.  
Algebraically, these include logarithms of celestial distances, but physically they are 
ultraviolet renormalisation-group logarithms of a bulk coupling. One could argue that 
the strict two-dimensional calculation may supply the angular kernel, but it does not
involve the energy $\omega_q$, the Laurent poles, or the QCD beta function: this is 
what happens for the celestial expression of the virtual factor $Z_n$. In other words,
while one may search for a (logarithmic) CCFT that dynamically reproduces this information
\cite{Bhardwaj:2024}, current QCD data do not require that interpretation.

Importantly, the scale organisation is not an accident of the one-loop formula.  At 
$L$ loops, dimensional analysis and independent rescaling invariance of every Wilson 
line assign to each dipole the ratio $(\mu^2/q_{ij}^2)^{L\eps}$, precisely the factor 
needed to form $\as^L(q_{ij}^2,\eps)$.  In particular, the complete two-loop result 
\cite{Dixon:2020} and the three-loop two-hard-leg current \cite{Herzog:2023} verify 
this pattern.  Finite constants are reshuffled between perturbative orders, but every 
logarithm generated by expanding the dipole scale factor is an ordinary bulk RG logarithm.  
Higher-loop multipoles contain additional cross-ratio-dependent angular functions; those 
are genuine celestial data, whereas the energy, regulator and scale dependences discussed 
here may be entirely assigned to the strong coupling.

%%%%%%%%%%%%%%%%%%%%%%%%%%%%%%%%%%%%%%

\subsection{Mixed-helicity multiple radiation and weighted celestial OPEs}
\label{subsec:multiple-soft}

The emission of two or more soft gluons probes products of celestial currents. In QCD, the 
leading-power factorisation formula defining soft currents is expected to apply for any 
multiplicity; explicit double-soft corrections are known at one loop~\cite{Zhu:2020ftr, 
Czakon:2023}, while the fully correlated triple-soft current is known at tree level 
\cite{Catani:2020}.  Collinear limits of emitted gluons are especially informative, 
because they isolate singular terms that should be reproduced by a celestial operator
product expansion (OPE). The order of limits is an important ingredient of this statement: 
for us, the soft limit is taken first, and uniformly. More precisely, for $m$ emitted gluons,
one scales $q_r \to \lambda q_r$ at fixed ratios $q_r/q_s$ and extracts
\beq
\label{eq:uniform-soft-limit}
  \mathcal A_{n+m} \big( \{\lambda q_r\}; \{p_i\} \big) \, = \, 
  \lambda^{-m} \, S^{(m)} \big(\{q_r\};\{p_i\} \big) \, \mathcal A_n \big(\{p_i\}\big)
  + \order \big(\lambda^{-m+1} \big) \, ;
\eeq
only then is a collinear limit taken inside the already defined soft current. A physical
collinear limit sends both $z_{12}$ and $\bar z_{12}$ to zero, so that the two real null
directions coincide.  This is distinct from a holomorphic collinear limit, in which $z$
and $\bar z$ are first treated as independent complex variables and only one separation
vanishes. Soft and collinear limits need not commute; the soft-first prescription is the
one relevant if celestial currents are meant to encode the universal long-wavelength
sector rather than the complete hard-amplitude.

For two positive-helicity gluons, the collinear limit $z_1\to z_2$ suggests a celestial
OPE
\beq
\label{eq:same-helicity-ope}
   j^{a_1}(z_1) \, j^{a_2}(z_2) \, \sim \, 
   \frac{\mathrm{i} f^{a_1a_2a}}{z_{12}} \, j^a(z_2) \, ,
\eeq
with an anti-holomorphic copy for two negative helicities. This is the familiar non-abelian 
Kac--Moody current algebra, without a tree-level central term. It also has a transparent 
bulk origin: when two emitted gluons become collinear, the dominant diagram contains one 
nearly on-shell parent gluon, and its splitting originates the structure constant and 
the simple pole.

Mixed helicities are more revealing.  Let $x_{ij} \equiv \omega_i/\omega_j$ be the ratio of 
two soft gluon energies. The QCD collinear limit requires both holomorphic and anti-holomorphic 
currents to contribute to the celestial OPE, according to
\beq
\label{eq:mixed-ope}
  j^{a_1}(z_1) \, \bar j^{a_2}(\bar z_2) \, \sim \, \mathrm{i} f^{a_1a_2a} \,
  \left[ \frac{1}{(1+x_{21}) \, \bar z_{12}} \, j^a(z_1)
  + \frac{1}{(1+x_{12}) \, z_{12}} \, \bar j^a(\bar z_2) \right] \, .
\eeq
This expression is unambiguous for any finite non-zero energy ratio, but it is very
unconventional from a two-dimensional viewpoint.  First, the OPE does not factor into
independent holomorphic and anti-holomorphic sectors.  Second, its coefficients remember
the four-dimensional energy fractions, bulk data that were so far not required to appear
in the celestial theory. The two strong orderings $\omega_1 \ll \omega_2$ and $\omega_2
\ll \omega_1$ select different terms in \eq{eq:mixed-ope}: in those limits, the expression
simplifies but it does not define a unique unordered boundary OPE. The energy dependence 
is especially transparent in terms of the energy fractions
\beq
\label{eq:soft-energy-fractions}
  y_1 \, = \, \frac{\omega_1}{\omega_1 + \omega_2} \, , \quad
  y_2 \, = \, \frac{\omega_2}{\omega_1 + \omega_2} \, ,  
  \quad \rightarrow \quad
  \frac{1}{1 + x_{21}} \, = \, y_1 \, , \quad \frac{1}{1 + x_{12}} \, = \, y_2 \, .
\eeq
Essentially, the uniform scaling in \eq{eq:uniform-soft-limit} removes the common soft 
energy, but leaves a dependence on the finite ratios $y_1$ and $y_2$. \eq{eq:mixed-ope} 
therefore gives a smooth, bulk-determined interpolation between the two strongly ordered 
endpoints.  From a two-dimensional viewpoint this means that position and colour labels 
are insufficient: the product of currents must retain an energy or boost-weight label if 
it is to describe the whole soft boundary of phase space.

At one loop, double emission also makes the scale-setting problem less constrained than for 
the single-soft current. With two soft momenta, there are several inequivalent transverse
scales available, even for a fixed hard dipole, as well as dimensionless ratios built 
from them.  The complete current consequently contains logarithms and dilogarithms 
of several kinematic ratios, and dimensional analysis alone cannot select a unique 
running-coupling argument analogous to $q_{ij}^2$. In strongly-ordered limits, the 
harder soft gluon behaves as an additional Wilson line sourcing the softer emission 
\cite{Magnea:2024jqg}.  The double current then reduces to nested single-emission 
structures and their one-loop corrections, recovering dipole-like scale assignments.  
This simplification is useful evidence for a celestial OPE, but it applies only to a
special hierarchical case, rather than to the full double-soft current.

Triple-soft radiation tests associativity by allowing two collinear limits to be taken in
succession.  For three equal helicities, the two orders of limits reproduce the colour Jacobi
identity and are consistent with repeated use of \eq{eq:same-helicity-ope}.  With mixed 
helicities, different orders combine different intermediate energy fractions.  The
two answers therefore disagree: the putative OPE is not associative in this sector, in
agreement with the general locality analysis of Ref.~\cite{Ball:2023}.  Complete strong
ordering restores an associative nested emission picture, because each softer gluon sees
the harder gluons as additional Wilson lines.  The result, however, still depends on the
chosen ordering. These bulk constraints suggest that a local, holomorphically factorised 
CFT of single-particle currents is too restrictive to encode generic non-abelian radiation.  
Possible extensions include operators retaining energy or boost-weight data, continuous 
energy-flow observables, or a celestial theory that is restricted to the strongly 
ordered soft regime.

%%%%%%%%%%%%%%%%%%%%%%%%%%%%%%%%%%%%%%%%%%%%%%%%%%%%%%%%%%%%%%%%%%%%%%%

\section{Outlook}
\label{sec:outlook}

We have summarised some recent progress in a remarkable line of research initiated by
Strominger more than a decade ago (see~\cite{Strominger:2017}), connecting the long-distance 
behaviour of $d=4$ quantum field theories, such as QED, gravity and Yang-Mills theories, with 
two-dimensional physics on the {\it celestial sphere}. This approach, {\it flat-space holography}, 
attempts to use the powerful tools of $d=2$ CFT to solve highly non-trivial problems 
in $d=4$ Minkowski space.

Our contribution has focused on non-abelian gauge theories, and we have deliberately taken 
a narrow view, aiming to find a holographic representation for the infrared factor of scattering 
amplitudes. We reviewed the construction of~\cite{Magnea:2021}, which gives a two-dimensional 
representation of colour-dipole contributions to the infrared factor, in terms of a free-boson 
theory, valid to all orders. This theory also correctly predicts the form of the soft emission 
current, albeit only at tree level. We then discussed the results of~\cite{Magnea:2026}, where
expressions for soft gluon currents at higher order were used to constrain scalar interactions 
on the sphere. Notably, {\it all} logarithms in the single-soft dipole current arise from renormalisation,
and are of ultraviolet origin. Furthermore, mixed-helicity multiple-emission currents require 
a form of energy-weighted OPE on the sphere, which breaks holomorphic factorisation, and 
eventually associativity for triple emissions. These results indicate that an interacting 
celestial description must retain more of the bulk radiative data than an ordinary local 
CFT of unlabelled currents. Further high-order corrections will no doubt continue to test 
the allowed color and kinematic structures which must be implemented on the sphere, hopefully 
delivering new understanding, and novel computational tools for as yet unknown corrections.

%%%%%%%%%%%%%%%%%%%%%%%%%%%%%%%%%%%%%%%%%%%%%%%%%%%%%%%%%%%%%%%%%%%%%%%

%%%%%%%%%%%%%%%%%%%%%%%%%%%%%%%%%%%%%%%%%%%%%%%%%%%%%%%%%%%%%%%%%%%%%%%


\begin{thebibliography}{99}

\bibitem{Kinoshita:1962ur}
T.~Kinoshita,
Mass singularities of Feynman amplitudes,
J. Math. Phys. \textbf{3} (1962), 650-677,
\url{https://doi.org/10.1063/1.1724268}.

\bibitem{Lee:1964is}
T.~D.~Lee and M.~Nauenberg,
Degenerate Systems and Mass Singularities,
Phys. Rev. \textbf{133} (1964), B1549-B1562,
\url{https://doi.org/10.1103/PhysRev.133.B1549}.

\bibitem{Agarwal:2023}
N. Agarwal, L. Magnea, C. Signorile-Signorile, A. Tripathi,
The infrared structure of perturbative gauge theories,
Phys. Rep. \textbf{994}, 1--120 (2023),
\url{https://doi.org/10.1016/j.physrep.2022.10.001}.

\bibitem{Strominger:2017}
A. Strominger, \textit{Lectures on the Infrared Structure of Gravity and Gauge Theory},
Princeton University Press, Princeton (2018),
\url{https://arxiv.org/abs/1703.05448}.

\bibitem{Strominger:2014}
A. Strominger, Asymptotic symmetries of Yang--Mills theory,
JHEP \textbf{07}, 151 (2014),
\url{https://doi.org/10.1007/JHEP07(2014)151}.

\bibitem{Arkani-Hamed:2020gyp}
N.~Arkani-Hamed, M.~Pate, A.~M.~Raclariu and A.~Strominger,
Celestial amplitudes from UV to IR,
JHEP \textbf{08} (2021), 062,
\url{https://doi.org/:10.1007/JHEP08(2021)062}.

\bibitem{Magnea:2021}
L. Magnea, Non-abelian infrared divergences on the celestial sphere,
JHEP \textbf{05}, 282 (2021),
\url{https://doi.org/10.1007/JHEP05(2021)282}.

\bibitem{Gardi:2009}
E. Gardi, L. Magnea, Factorization constraints for soft anomalous dimensions in QCD
scattering amplitudes,
JHEP \textbf{03}, 079 (2009),
\url{https://doi.org/10.1088/1126-6708/2009/03/079}.

\bibitem{Becher:2009}
T. Becher, M. Neubert, Infrared singularities of scattering amplitudes in perturbative QCD,
Phys. Rev. Lett. \textbf{102}, 162001 (2009),
\url{https://doi.org/10.1103/PhysRevLett.102.162001}.

\bibitem{Almelid:2016}
\O. Almelid, C. Duhr, E. Gardi, Three-loop corrections to the soft anomalous dimension in
multileg scattering,
Phys. Rev. Lett. \textbf{117}, 172002 (2016),
\url{https://doi.org/10.1103/PhysRevLett.117.172002}.

\bibitem{Catani:2019}
S. Catani, D. de Florian, M. Grazzini, Soft-gluon effective coupling and cusp anomalous
dimension,
Eur. Phys. J. C \textbf{79}, 685 (2019),
\url{https://doi.org/10.1140/epjc/s10052-019-7174-9}.

\bibitem{He:2016}
T. He, P. Mitra, A. Strominger, 2D Kac--Moody symmetry of 4D Yang--Mills theory,
JHEP \textbf{10}, 137 (2016),
\url{https://doi.org/10.1007/JHEP10(2016)137}.

\bibitem{Magnea:2026}
L. Magnea, E. Zunino, Non-abelian soft radiation data for a celestial theory,
Class. Quantum Grav. \textbf{43}, 125007 (2026),
\url{https://doi.org/10.1088/1361-6382/ae7794}.

\bibitem{Catani:2000}
S. Catani, M. Grazzini, The soft-gluon current at one-loop order,
Nucl. Phys. B \textbf{591}, 435--454 (2000),
\url{https://doi.org/10.1016/S0550-3213(00)00572-1}.

\bibitem{Bhardwaj:2024}
R. Bhardwaj, A. Yelleshpur Srikant, Celestial soft currents at one-loop and their OPEs,
JHEP \textbf{07}, 034 (2024),
\url{https://doi.org/10.1007/JHEP07(2024)034}.

\bibitem{Dixon:2020}
L.J. Dixon, E. Herrmann, K. Yan, H.X. Zhu, Soft gluon emission at two loops in full color,
JHEP \textbf{05}, 135 (2020),
\url{https://doi.org/10.1007/JHEP05(2020)135}.

\bibitem{Herzog:2023}
F. Herzog, Y. Ma, B. Mistlberger, A. Suresh,
Single-soft emissions for amplitudes with two colored particles at three loops,
JHEP \textbf{12}, 023 (2023),
\url{https://doi.org/10.1007/JHEP12(2023)023}.

\bibitem{Zhu:2020ftr}
Y.~J.~Zhu, Double soft current at one-loop in QCD,
JHEP \textbf{04}, 018 (2026)
\url{https://doi.org/10.1007/JHEP02(2026)018}.

\bibitem{Czakon:2023}
M. Czakon, F. Eschment, T. Schellenberger,
Revisiting the double-soft asymptotics of one-loop amplitudes in massless QCD,
JHEP \textbf{04}, 065 (2023),
\url{https://doi.org/10.1007/JHEP04(2023)065}.

\bibitem{Catani:2020}
S. Catani, D. Colferai, A. Torrini,
Triple (and quadruple) soft-gluon radiation in QCD hard scattering,
JHEP \textbf{01}, 118 (2020),
\url{https://doi.org/10.1007/JHEP01(2020)118}.

\bibitem{Magnea:2024jqg}
L.~Magnea, C.~Milloy, C.~Signorile-Signorile and P.~Torrielli,
Strongly-ordered infrared counterterms from factorisation,
JHEP \textbf{06} (2024), 021, 
\url{https://doi.org/10.1007/JHEP06(2024)021}.

\bibitem{Ball:2023}
A. Ball, Celestial locality and the Jacobi identity,
JHEP \textbf{01}, 146 (2023),
\url{https://doi.org/10.1007/JHEP01(2023)146}.

\end{thebibliography}
\end{document}